# Flat Electronic Bands in Long Sequences of Rhombohedral-stacked Multilayer Graphene

Hugo Henck[1], Jose Avila[2], Zeineb Ben Aziza[1], Debora Pierucci[3], Jacopo Baima[4], Betül Pamuk[4], Julien Chaste[1], Daniel Utt[5], Miroslav Bartos[5], Karol Nogajewski[5], Benjamin A. Piot[5], Milan Orlita[5], Marek Potemski[5], Matteo Calandra[4], Maria C. Asensio[2], Francesco Mauri[6,7], Clément Faugeras[5,*] and Abdelkarim Ouerghi[1,*]

[1]Centre de Nanosciences et de Nanotechnologies, CNRS, Univ. Paris-Sud, Université Paris-Saclay, C2N – Marcoussis, 91460 Marcoussis, France

[2] Synchrotron-SOLEIL, Saint-Aubin, BP48, F91192 Gif sur Yvette Cedex, France

[3]CELLS - ALBA Synchrotron Radiation Facility, Carrer de la Llum 2-26, 08290 Cerdanyola del Valles, Barcelona, Spain

[4] Institut de Minéralogie, de Physique des Matériaux, et de Cosmochimie, UMR CNRS 7590, Sorbonne Universités, UPMC, Univ. Paris VI, MNHN, IRD, 4 Place Jussieu, 75005 Paris, France

[5] Laboratoire National des Champs Magnétiques Intenses, CNRS-UGA-UPS-INSA-EMFL, 25 rue des Martyrs, 38042 Grenoble, France

[6] Departimento di Fisica, Università di Roma La Sapienza, Piazzale Aldo Moro 5, I-00185 Roma, Italy

[7]Graphene Laboratories, Fondazione Istituto Italiano di Tecnologia, 16163 Genova, Italy

*Corresponding author, E-mail: abdelkarim.ouerghi@c2n.upsaclay.fr
clement.faugeras@lncmi.cnrs.fr

The crystallographic stacking order in multilayer graphene plays an important role in determining its electronic properties. It has been predicted that a rhombohedral (ABC) stacking displays a conducting surface state with flat electronic dispersion. In such a flat band, the role of electron-electron correlation is enhanced possibly resulting in high $T_c$ superconductivity, charge density wave or magnetic orders. Clean experimental band structure measurements of ABC stacked specimens are missing because the samples are usually too small in size. Here, we directly image the band structure of large multilayer graphene flake containing approximately 14 consecutive ABC layers. Angle-resolved photoemission spectroscopy experiments reveal the flat electronic bands near the K point extends by 0.13 Å$^{-1}$ at the Fermi level at liquid nitrogen temperature. First-principle calculations identify the electronic ground state as an antiferromagnetic state with a band gap of about 40 meV.

Multilayer graphene materials present two different polytypes describing the way individual monolayer are stacked one on top of another[1]. The stable polytype of multilayer graphene is the so-called Bernal stacking, with an ABA sequence of the graphene layers. Rhombohedral multilayer graphene, with an ABC sequence, is less stable in ambient conditions and hence it is less abundant[2]. With all the recent efforts to (re)visit the electronic properties of multilayer graphene and of thin layers of graphite, ABC stacked trilayers have been successfully identified and studied experimentally[3]. Rhombohedral trilayer graphene shows many remarkable properties such as a band gap at low energy with a magnitude tunable by applying an electric field[3,4,5] and it also hosts chiral massive Dirac fermions as evidenced by their peculiar Quantum Hall Effect[6,7]. Transport measurements on suspended ABC trilayers display the opening of a 40 meV gap, tentatively attributed to a magnetic state[8]. However, this magnetic state is fragile as it occurs only in ultra-clean samples and is destroyed by temperature ($T_c$= 30 - 40 K) or by the presence of a substrate.

Rhombohedral multilayer graphene, ABC-stacked layers of graphene, is predicted to host an electronic band with a flat dispersion at the K points of the Brillouin zone. This system shares many common features with crystalline topological insulators[9]. The associated wave functions are spatially located on its top and bottom surfaces, while bulk states present an energy gap[10]. In the following, we will refer to these bands as the flat bands or as surface states. When increasing the number of ABC stacked layers, the extent in k-space of the flat dispersion increases and the energy gap in the bulk decreases[1]. The electronic ground state in a flat band has been predicted to be superconducting[11,12] or magnetic[6,13,14,15]. Experiments performed on mixtures of AB and ABC stacked multilayer graphene are interpreted in terms of superconductivity[16,12,17,18] but there exist up to now no experiments addressing macroscopic single crystals with a large ABC stacked sequence mainly due to the lack of established ABC-stacked multilayer graphene source. Thick specimens are seldom because this polytype is less abundant than the stable Bernal stacked multilayer graphene and their electronic properties remain unknown. A diverging density of states, characteristic of a gapless electronic band with a flat dispersion has been observed in scanning tunneling spectroscopy in 5 graphene layers with ABC stacking grown on 3C-SiC[19]. A larger sequence of 15 ABC-stacked graphene layers has recently been produced by exfoliation, and its electronic band structure was determined by Landau level spectroscopy[20]. A proper and systematic investigation of the electronic properties of long-sequences of ABC stacked multilayer graphene is still lacking.

In this work, we use angular resolved photoemission spectroscopy (ARPES) to investigate the band structure of a large sequence of ABC stacked graphene layers produced by exfoliation. We report on the observation of the band structure of thin flake rhombohedral multilayer graphene, in particular the low energy bands at the K point displaying a flat dispersion. This band structure is compared to the one of a Bernal stacked domain of similar thickness. We underline that the dispersion of the surfaces states band is about 25 meV and the flattening of the band top is approximately 0.13 A$^{-1}$ near the K points. Ab initio calculations identify this flat band dispersion as originating from the exchange interaction among carriers in the half-filled bands on the two surfaces. The electronic ground state is hence spin polarized and thin layers of rhombohedral multilayer graphene can offer an interesting platform for spintronics applications.

Graphene multilayers exhibit two structures with Bernal (Figure 1(a)) and rhombohedral (Figure 1(b)) stacking (ABA and ABC respectively). Our thin rhombohedral multilayer graphene layer was produced by mechanical exfoliation of Kish graphite on a polydimethylsiloxane (PDMS) stamp. Flakes with ABC stacking were then identified based on their Raman scattering response[20,21] and, in particular, on the line shape of the 2D band feature. Because the 2D band feature arises from a double resonant Raman scattering process, it directly involves the electronic band structure and is hence a sensitive probe of the stacking configuration. The 2D band in ABC stacked multilayer graphene appears as a multi-component feature which is significantly broader than the one in ABA stacked thin multilayer graphene. Moreover, the occurrence of a shoulder at 2576 cm$^{-1}$ at 1.96 eV laser light is a criterion to discriminate between the two stacking configurations[20,21](Figure 1(c) and (d)). Clearly two domains are present in the flake: one with ABA stacking and the other with ABC stacking. The ABC part has a large spatial extension with size up to 20 x 80 micrometres (Figure 1(e)). More details on Raman scattering characterization of our sample are provided in the Supplementary information (SI). The selected flake was then deterministically transferred onto a graphitized 4H-SiC substrate, with a conductive graphene monolayer (1 ML Gr) at the surface, which is particularly well adapted for ARPES measurements[22]. The graphene used was obtained by annealing 4H-SiC(0001) at 1550 °C in 800 mbar argon for 10 min[23,24]. Figure 1(f) shows an optical micrograph of a rhombohedral multilayer graphene flake after its transfer to the graphitized 4H-SiC. Typical AFM images of the sample are reported in Figure S1(a). The different profile scan presented in Figure S1(b) allowed us to measure the height of the flake and we obtained an estimation of the layers number of about 50 – 60 Layers.

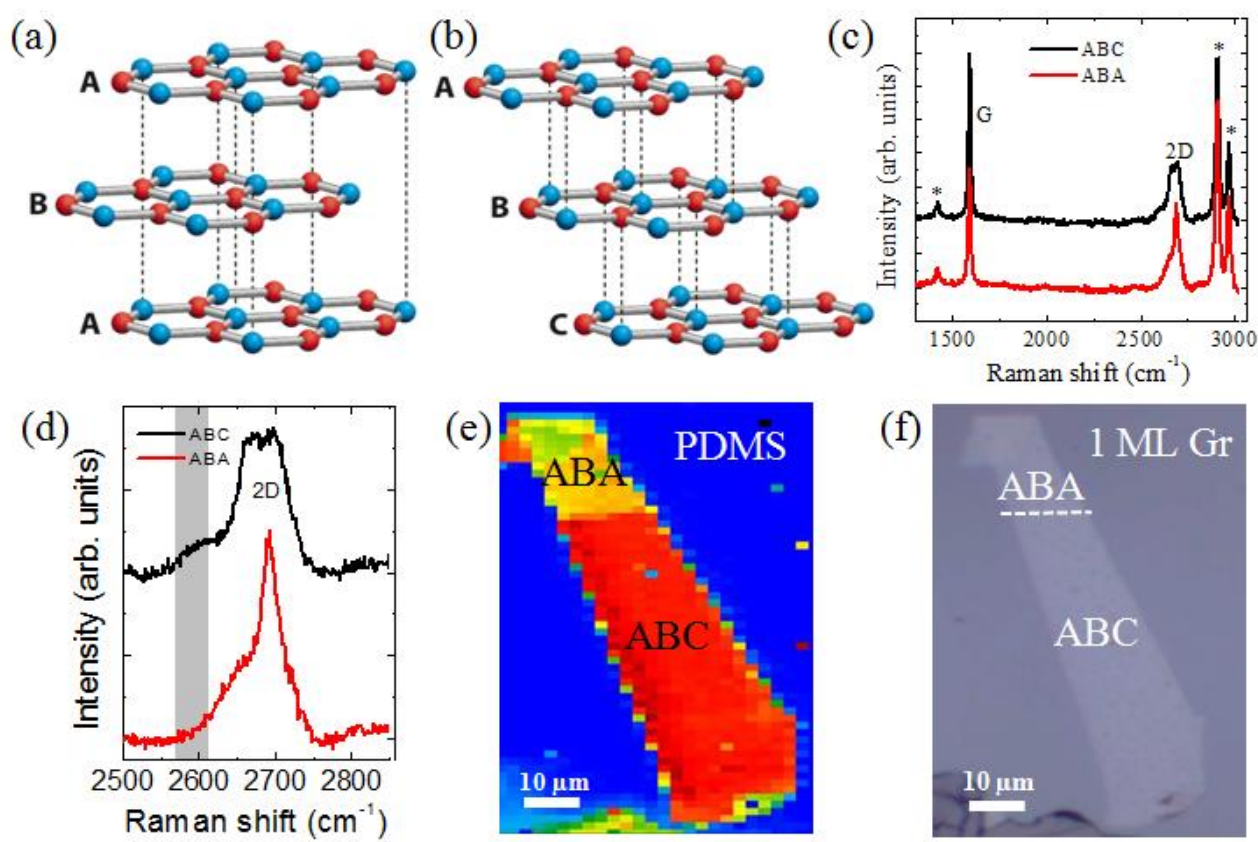

**Figure 1 (Color online): Schematic representation of rhombohedral multilayer graphene structure and their Raman signature: Top view of (a) ABA and (b) ABC stacking sequences of rhombohedral multilayer graphene. The colored layers differentiate the successive position in the in-plane direction were each color corresponds to an inequivalent layer position. (c) Raman scattering spectra of ABC- (black curve) and ABA- (red curve) stacked thin multilayer graphene measured on a flake on polydimethylsiloxane (PDMS) stamp.**

**The G band and 2D bands are indicated and the star symbols indicate Raman scattering features from PDMS. (d) Zoom of ABC- (black curve) and ABA- (red curve) stacked graphene 2D peak. (e) False color spatial map of the integrated intensity in the shaded energy range in Fig 1d, where a significant difference is observed between ABC and ABA, (f) Optical image of the studied flake.**

To resolve the electronic structure of rhombohedral multilayer graphene film described above, we have performed nano-ARPES measurements at synchrotron Soleil at liquid nitrogen temperature (90 K), with a spatial resolution of 100 nm[25]. In Figure 2(a) we show the averaged ARPES map coming from the rhombohedral multilayer graphene flake shown in figure 1(f). The photoemission intensity around the K-point for the rhombohedral multilayer graphene was acquired along the ΓKM direction for a photon energy hν = 100 eV as a function of space coordinate. The ARPES intensity was then integrated over different energy ranges illustrated by the three colored rectangles A, B and C in Figure 2(a). Figure 2(b) shows a spatially resolved map integrated from 1 eV to the Fermi energy $E_F$ (green rectangle) and demonstrates the homogeneity of the ARPES intensity over the flake, thus confirming its uniform thickness. Figure 2(c) and (d) show spatially-resolved maps integrated from 150 meV to 250 meV and from 100 meV to $E_F$, respectively (red and yellow rectangles respectively). Different regions on the flake can be identified based on these integrated intensity maps. The large background region with low ARPES intensity (outside the white dotted line) represents the band structure of the monolayer graphene/SiC substrate as labeled "1ML Gr" in Figure 1(f). As can be seen by the different ARPES intensities, the flake is composed of two areas of size in the μm range. In agreement with the micro-Raman measurement (figure 1(d)), these two area correspond to an ABA (figure 2(c)) and ABC (figure 2(d)) stacked multilayer graphene. The fact that ABA and ABC stacking present a different ARPES intensity in different energy ranges is strictly related to their different electronic structure as shown in figure 3.

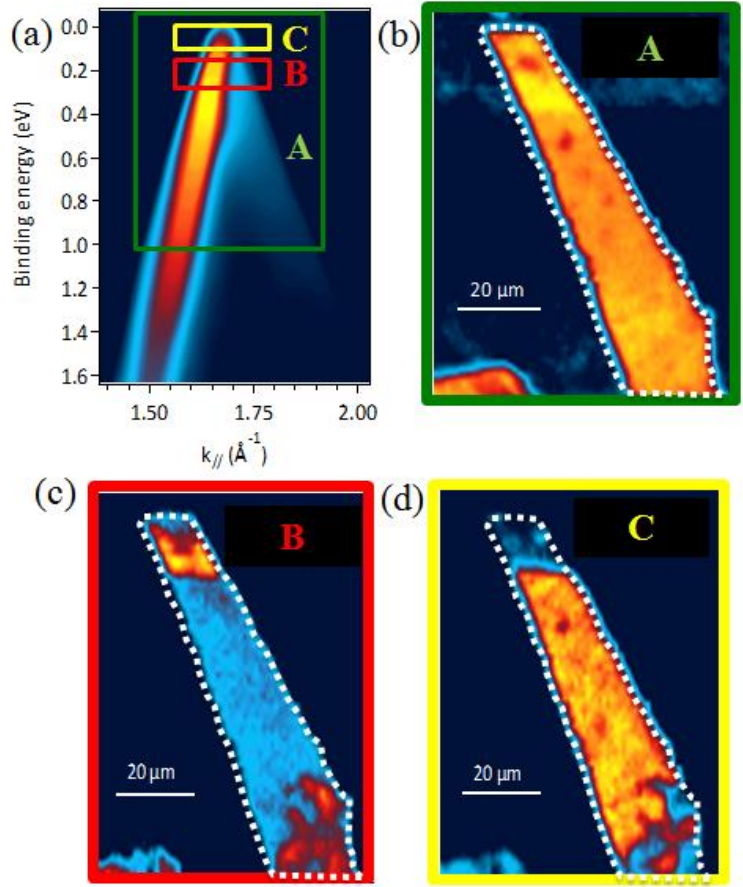

**Figure 2 (Color online): Spatially-resolved electronic structure of an exfoliated rhombohedral multilayer graphene flake: (a)** Nano-ARPES intensity map averaged over the entire flake of exfoliated multilayer graphene structure transferred onto graphene-SiC(0001) substrate . **(b-d)** Spatial false color map built from nanoARPES intensity map of (a) showing the homogeneity of the flake (b) and the ABA and ABC stacked inclusion (c-d). The colored rectangles in (a) indicate the energy integration range where the spatially resolved maps of (b-d) have been obtained.

Figure 3(a-c) show nano-ARPES intensity maps measured at three different locations: 1ML Gr, ABA and ABC that were identified by ARPES and micro-Raman scattering (see Figure 1(d) and Figure 2). Due to the transfer process, the BZ of rhombohedral multilayer graphene is rotated by 17° with respect to the graphene BZ (Figure 3(a)). The observed electronic band structures for the ABA and ABC areas are very different despite the similar thickness of the flake at these locations. Detailed dispersions of different π bands in the vicinity of the K and K' special points are well resolved. The linear dispersion of single layer graphene of the substrate below the flake is very clear in Figure 3(b). Figure 3(c) shows the band structure at the ABA location, where different bands can be observed. The main two bands have a quadratic dispersion, are split by ~200 meV and are identified as the two split-off bands close to the K point of the ABA multilayer graphene [26]. The ARPES response measured over the ABC domain is completely different (Figure 3(d)). It consists of an intense band close to $E_F$ with a flat dispersion, and two cone-like structures composed of a manifold of bands (Figure 3(e)). The flat band at the Fermi level is also confirmed by the energy distribution curves (EDC) at the K point with about 50 meV bandwidth (Figure 3(f)). This picture precisely reflects the expected band structure of rhombohedral multilayer graphene [13,15].

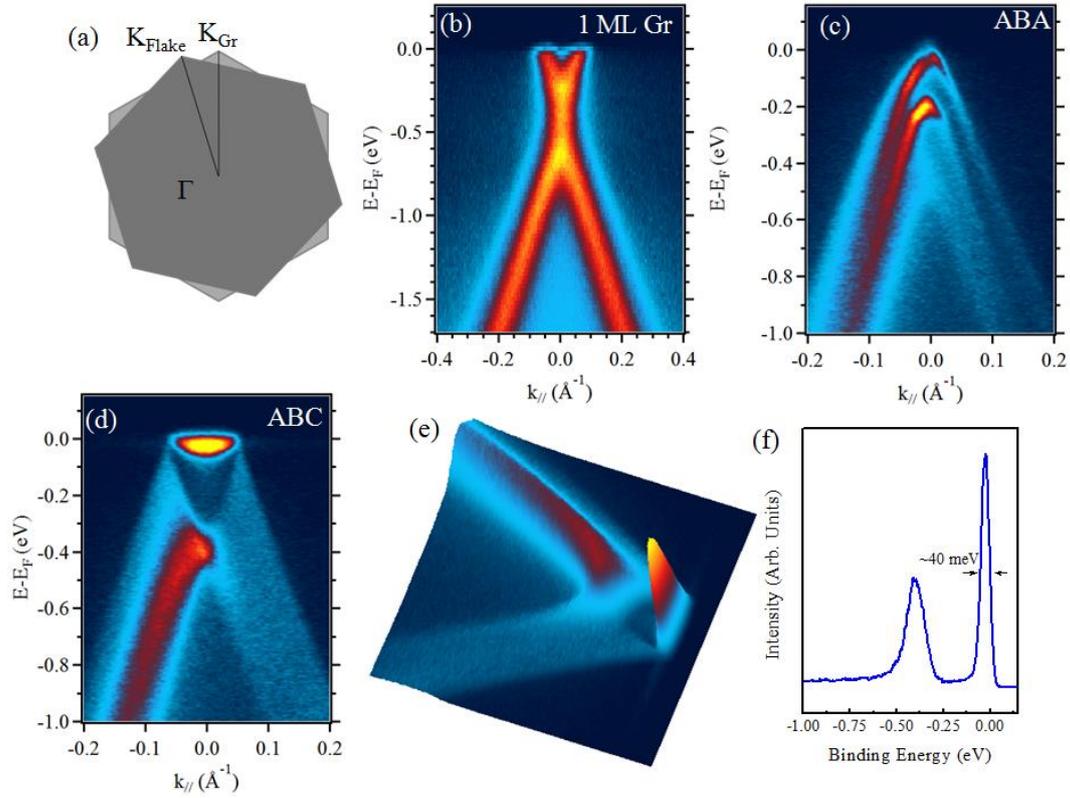

**Figure 3 (Color online): Angle-resolved electronic structure of an exfoliated rhombohedral multilayer graphene: a) Brillouin zone of graphene/SiC(0001) and thin stacked multilayer graphene flake with a twist angle of 17°, ARPES intensity maps around the K point of the Brillouin zone for (b) the monolayer graphene substrate, recorded in the direction perpendicular to ΓK, (c) ABA and (d) ABC regions of the flake recorded along the ΓKM direction. The spectra were measured with photon energy of 100 eV. (e) 3D plot of the ARPES intensity map for the ABC region shown in (d). (f) Energy distribution curve (EDC) of thin rhombohedrally stacked multilayer graphene film of (d) at the K point of the graphene Brillouin zone.**

From these results we can confirm that the sample is composed of a small domain with Bernal stacking and a larger domain with a broad ABC-stacked sequence of graphene layers, in accordance with Raman spectroscopy and a flat dispersion of the surface state. Following the maximum intensity in this map at energies well below the Fermi level, a linear-shaped dispersion is observed. One can also note that this sample appears to be quasi neutral, with the Fermi energy close the charge neutrality point (figure 3(f)). A more detailed inspection of the electronic band dispersion close to $E_F$ at the ABC location shows that it is not exactly flat, in fact a curvature in the ΓK and KM directions can be observed. From the analysis of the energy distribution curves (EDC) shown in Figure 4(a), we conclude that a higher density of surface states occurs near the Fermi level. The flat band near the K point extends by about 0.13 Å$^{-1}$, disperses in the ΓK and KM directions towards the valence band maximum, and extends over 25 meV. This can be particularly appreciated in the second derivative of the ARPES response with respect to energy shown in Figure 4(b) and (c). The flat band extension in k-space is strictly related to the length of the ABC-stacked sequence[27] corresponding, in our case, to 14 ABC-stacked graphene layers. **This value is different with the AFM measurement** and this difference can be attributed to a stacking fault (twisted angle of about 2°/3°) between the topmost and bottom layers, thus decoupling the 14 rhombohedral stacking layers from the bottom[19].

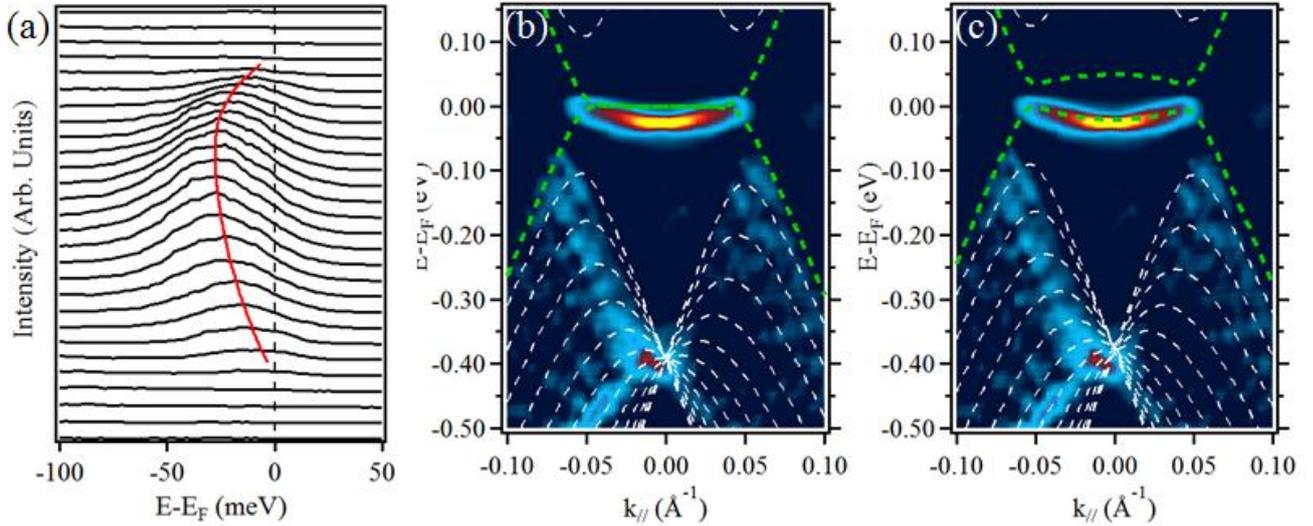

**Figure 4 (Color online): Energy vs momentum dispersion relation of rhombohedrally stacked multilayer graphene film flat band: (a) ARPES energy distribution curves of 14-layer sequence of ABC-stacked rhombohedral multilayer graphene showing the presence of a flat band around the charge neutrality point. (b) and (c) Second derivative of trhombohedrally stacked multilayer graphene film ARPES intensity map compared with theoretical calculation of 14-layer sequence (b) without and (c) with the spin polarization. White dashed curves in these graphs correspond to bulk states and green dashed curves highlight the surface state of the outer layers.**

To gain further insight into the electronic structure thin rhombohedrally stacked multilayer graphene film, we performed first-principles electronic structure calculations with inclusion of an exact exchange (see SI). We considered a 14-layer crystal with the ABC stacking, as indicated by the experimentally observed extend in k-space of the flat band of 0.13 Å$^{-1}$, and we included a 37 Å vacuum gap between the periodic images. We performed spin-polarized and unpolarized calculations. The same rescaling of the Fermi velocity found for graphene was applied. By neglecting spin polarization, the electronic structure displays an extremely narrow metallic surface state (band dispersion smaller than 2 meV) localized around the K-point in the Brillouin zone (see green dashed lines in Figure 4 (b)). This is in disagreement with the experiment, which established a substantially larger bandwidth (25 meV). The two cones formed by an ensemble of electronic bands corresponding to the bulk states, and intersecting at the K point at an energy of - 400 meV, are well reproduced. By introducing the spin polarization, an antiferromagnetic state is stabilized, in agreement with the findings of Ref.[15][15] obtained, however, for thinner flakes. The largest magnetic moments are on the outer layers and they gradually decrease towards the center of the flakes. The coupling between the layers is antiferromagnetic, while the coupling inside the layers is slightly ferrimagnetic. The magnetism does not affect the bulk states electronic structure but it results, as determined by the calculations, in an insulating surface state with a gap opening of 40 meV. Our calculations show that this band gap originates from the exchange interaction which lifts the spin degeneracy of the flat band states. Interestingly, the effective mass of the top of the valence band (surface state) is now in agreement with the 25 meV dispersion found in ARPES experiments (see Figure 4 (b) and (c)). The extension of the flat surface state in the Brillouin zone (0.13 Å$^{-1}$) is also well reproduced. The measured band is in excellent agreement with electronic structure calculations. Electronic structure calculations do indeed converge to a gapped magnetic state[15] with a Néel temperature of approximately 120K, energetically favorable with respect to the non magnetic state or to CDW states with √3x√3 periodicity. This plus the fact that transport measurements in trilayer ABC graphene[8] shows the occurrence of a 40 meV Gap, in agreement with the calculations of Pamuk et al[15], lead us to infer the possible occurrence of a magnetic gap in the measured spectrum. . Remarkably, while a Curie temperature of 34 K was determined from transport measurements done on ABC trilayers [8] here, in our thicker flake, the state appears to be stable at least up to liquid nitrogen temperature. However, magnetic or transport measurements are needed to further confirm this claim.

In summary, we have successfully isolated a thin rhombohedrally stacked multilayer graphene film with a long sequence of ABC stacked layers (14 layers). By performing high-resolution nano-ARPES experiments, we have spatially mapped its electronic band structure. Nano-ARPES experiments show the existence of a flat band with 25 meV dispersion and a large extension in k-space of 0.13 Å$^{-1}$. Our DFT calculations predict that the dispersion of the flat band is compatible with a magnetic ground state characterized by an energy band gap close to 40 meV.

**Acknowledgements:** This work was supported by the ANR H2DH and ANR-17-CE24-0030 grants, by the European Research Council (MOMB project no. 320590) and by the EC Graphene Flagship project (no. 604391). We acknowledge the support from the European Union Horizon 2020 research and innovation program under Grant agreement No. 696656-GrapheneCore1. B. P. acknowledges the support from the National Science Foundation (Platform for the Accelerated Realization, Analysis, and Discovery of Interface Materials (PARADIM)) under Cooperative Agreement No. DMR-1539918. Computer facilities were provided by CINES, IDRIS, and CEA TGCC (Grant EDARI No. 2017091202).

# Supplementary Information

# Flat Electronic Bands in Long Sequences of Rhombohedral-stacked Multilayer Graphene

Hugo Henck[1], Jose Avila[2], Zeineb Ben Aziza[1], Debora Pierucci[3], Jacopo Baima[4], Betül Pamuk[4], Julien Chaste[1], Daniel Utt[5], Miroslav Bartos[5], Karol Nogajewski[5], Benjamin A. Piot[5], Milan Orlita[5], Marek Potemski[5], Matteo Calandra[4], Maria C. Asensio[2], Francesco Mauri[6,7], Clément Faugeras[5,*] and Abdelkarim Ouerghi[1,*]

[1]Centre de Nanosciences et de Nanotechnologies, CNRS, Univ. Paris-Sud, Université Paris-Saclay, C2N – Marcoussis, 91460 Marcoussis, France

[2] Synchrotron-SOLEIL, Saint-Aubin, BP48, F91192 Gif sur Yvette Cedex, France

[3]CELLS - ALBA Synchrotron Radiation Facility, Carrer de la Llum 2-26, 08290 Cerdanyola del Valles, Barcelona, Spain

[4] Institut de Minéralogie, de Physique des Matériaux, et de Cosmochimie, UMR CNRS 7590, Sorbonne Universités, UPMC, Univ. Paris VI, MNHN, IRD, 4 Place Jussieu, 75005 Paris, France

[5] Laboratoire National des Champs Magnétiques Intenses, CNRS-UGA-UPS-INSA-EMFL, 25 rue des Martyrs, 38042 Grenoble, France

[6] Departimento di Fisica, Università di Roma La Sapienza, Piazzale Aldo Moro 5, I-00185 Roma, Italy

[7]Graphene Laboratories, Fondazione Istituto Italiano di Tecnologia, 16163 Genova, Italy

**Methods:** The 4H-SiC(0001) substrate was etched with hydrogen (100% $H_2$) at 1550°C for 10 min. Subsequently, the substrate was annealed at 1000°C under $10^{-6}$ mbar of argon and then at 1550°C under 800 mbar of argon for 10 min[1]. To produce thin rhombohedral multilayer graphene, we have exfoliated Kish graphite onto a PDMS stamp. The flakes of interest were then deterministically transferred onto a graphitized 4H-SiC substrate, suitable for ARPES measurements. The quality of the multilayer graphene transferred onto the graphene underlayer was proved by atomic force microscopy (AFM) and Raman spectroscopy measurements.

**ARPES measurement:** The sample was annealed at 800 °C to clean the surface before the ARPES measurements. Nano-ARPES experiments were performed on the ANTARES beamline (SOLEIL French synchrotron facility). The ARPES data were taken at photon energy of 100 eV with Scienta R4000 analyzer, using linearly polarized light at a base pressure of $5 \times 10^{-11}$ mbar. The sample was kept at 90 K during the ARPES measurement. The sample orientation was set as to explore the k-space region around the K point in the ΓKM direction of the Brillouin zone.

**First-principle electronic structure calculations:** We used the PBE0 functional as implemented in the CRYSTAL code[2] and a triple-ζ -polarized Gaussian-type basis set[3]. All technical details are the same as in. Ref. [4]. We first performed PBE0 calculations on graphene and determined that a 16% reduction of the Fermi velocity is needed to match the experimental electronic structure.[5]

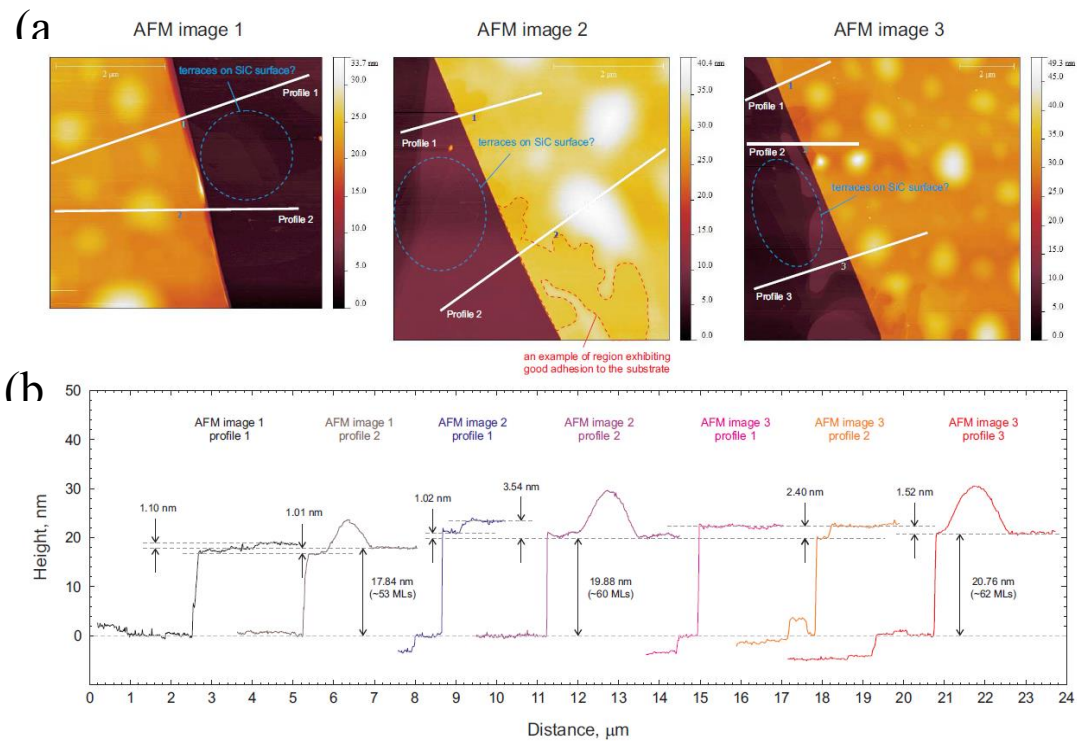

**Figure S1: AFM images of ABC stacking graphene multilayer**: a) AFM image of graphene multilayer on epitaxial graphene, b) different AFM height profiles corresponding to the line in (a).

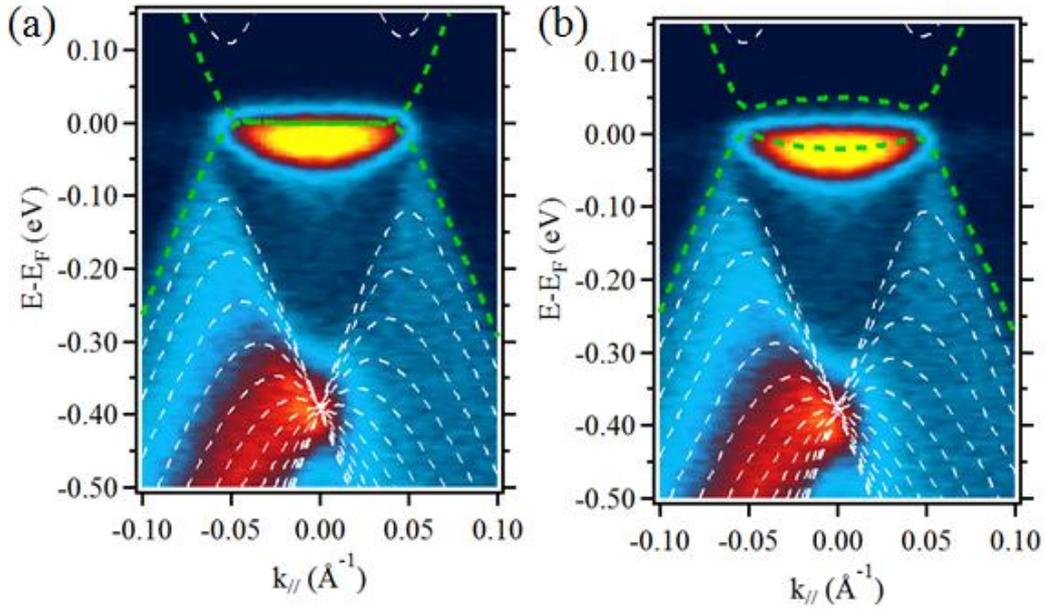

**Figure S2: Energy vs momentum dispersion relation of ABC-stacked graphene multilayer's flat band:** (a) and (b) ARPES intensity map of ABC graphene compared with theoretical calculation of 14-layer sequence of multilayer graphene (a) without and (b) with the spin polarization. White dashed curves in these graphs correspond to bulk states and green dashed curves highlight the surface state of the outer layers.

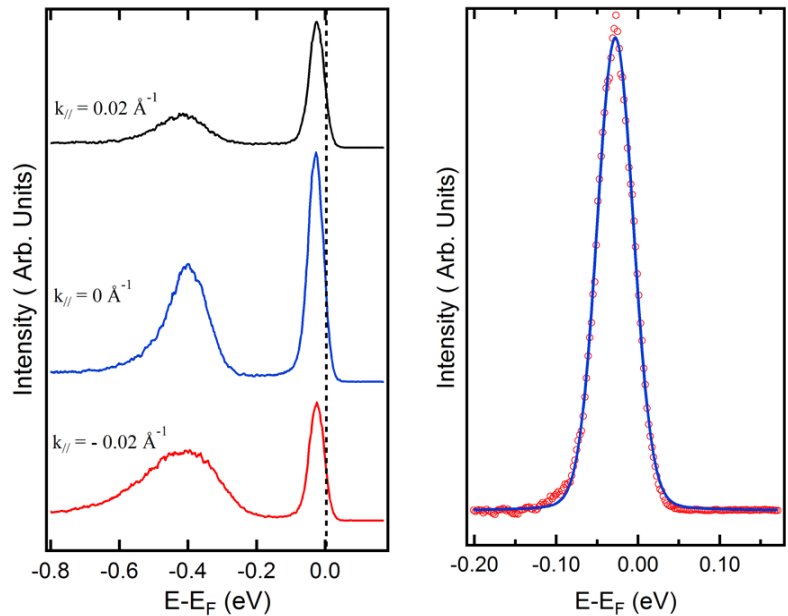

**Figure S3: Energy distribution curves (EDCs):** (a) EDC curves extracted from the ARPES map of figure 3(d) at $k_{//}$ = -0.02 Å$^{-1}$, 0 Å$^{-1}$ and 0.02 Å$^{-1}$. (b) Voigt line shape fitting of the EDC curve (FWHM = 0.05 eV)